# Exotic centrosymmetric phase of acentric urea under high pressure

Haw-Tyng Huang, Yedukondalu Neelam, Mei-Shuan Cheng, Zhenxian Liu, Lkhamsuren Bayarjargal, Rachel Husband, Anna Pakhomova, John B. Parise, Lars Ehm

Mineral Physics Institute, Stony Brook University, Stony Brook , New York 11790, USA.

*Corresponding author(s). E-mail(s): hawtynghuang@ntu.edu.tw; nykondalu@gmail.com; lars.ehm@stonybrook.edu;

**Abstract**

Urea is a simple prototype supramolecular crystal that exhibits rich polymorphism at low pressure due to broken and restored N-H···O hydrogen bonds. The high pressure polymorph (phase $\boldsymbol{V'}$) of acentric urea crystallizes in a centrosymmetric structure, which presents an appealing target because of its potential exotic structure, analogous to the symmetric ice phase X. The pressure-induced polymorphism of urea was studied using powder X-ray diffraction, infrared and Raman spectroscopy, second harmonic generation (SHG) measurements up to 20 GPa and *ab initio* crystal structure prediction (CSP) based on the constrained evolutionary approach. A strong decrease of the SHG signal at the transition pressure 10 GPa reveals that the high-pressure polymorph is indeed centrosymmetric, further confirmed by the selection rules observed in the lattice vibration modes, in contrast to chemical intuition for acentric urea. The structural evolution sequence obtained from X-ray diffraction, SHG and CSP calculations is as follows: phase I ($\boldsymbol{P\bar{4}2_1m}$; Z=2) from 0 to 0.5 GPa, Phase III ($P2_12_12_1$; Z=4) from 0.5 to 5.2 GPa, and phase $\boldsymbol{V'}$ ($P2_1/m$; Z=6) beyond 10.0 GPa which is energetically competitive with the theoretically predicted phase V ($\boldsymbol{Pnma}$; Z=4). A phase X with distinct spectral and diffraction features forms between 5.2 and 10.0 GPa, which could be explained by a quantum disorder intermediate state between phase III and $\boldsymbol{V'}$, that is ascribed to the difficulty to disrupt the H-bonding network under extremely compressed environment. The softening of N-H vibrations and the change in intensity of the vibrations associated with the hydrogen bonding provide evidence for proton tunneling and charge-transfer interaction in phase X.

# 1 Introduction

Urea is a simple supramolecular crystal and it was the first organic compound which was synthesized in laboratory from inorganic precursors in 1828 and also the first organic crystal studied by using X-ray diffraction (XRD) technique in 1921.[1] Urea finds potential applications as a fertilizer due to its high nitrogen content and serves as a precursor for pharmaceutical and cosmetic industries. In addition, it can also serve as hydrogen storage and transport material due to its high hydrogen content (7.95 wt.%), eco-friendly, odourless and non-flammable nature.[2] Despite of its potential applications, urea also received fundamental interest due to its rich polymorphism at very low pressure and temperature (P-T) conditions.[3, 4] Therefore, extensive experimental and theoretical studies were reported in the literature by addressing its polymorphs and hydrogen bond breaking/rearrangements using X-ray diffraction, (IR and Raman) spectroscopic measurements and density functional theory calculations.[5]

The previous neutron diffraction study[6] of urea under pressure suggests that urea crystallizes in phase I ($P\bar{4}2_1m$, Z = 2) tetragonal crystal symmetry at ambient conditions and transforms into an orthorhombic phase III ($P2_12_12_1$, Z = 4) at 0.48 GPa and later phase III transforms to phase IV ($P2_12_12$, Z = 2) above 2.8 GPa. Upon further compression, phase IV transforms to an intriguing centrosymmetric phase V ($Pnma$, Z = 4) above 7.2 GPa based on neutron diffraction measurements.[6] However, a similar study on hydrogenated urea finds the phase changes take place at pressure near 0.5 GPa, 5.0 GPa, and 8.0 GPa instead.[7] Furthermore, the single-crystal Raman mapping study (Lamelas et al.) observed spatial disorder, resulting in two co-existing distinct structural domains within the single crystal. [7] This indicates that the ordered models might not be satisfactory for the high pressure phases of urea. These reported differences and discrepancies in the phase transition pressures and space group symmetry between $CO(NH_2)_2$ and $CO(ND_2)_2$ at high pressures have important implications. First, deuterated compounds are widely used for neutron diffraction studies to avoid large incoherent scattering from hydrogen. However, the assumption often made implicitly that substituting hydrogen with deuterium will not affect the structures may not be valid. For example, the H-D substitution in brucite ($Mg(OH)_2$) has a significant effect on the unit-cell volume, which is attributed to the preferential incorporation of deuterium over hydrogen under pressure.[8] [9] Therefore, the deuterated analogs to hydogenous compounds should be used with caution, especially in hydrogon-bonding networks. More importantly, such subtle differences in the structure between hydrogenated and dueterated compounds can provide important insight into hydrogen bonds under pressures such as the exotic phenomena associated with proton transfer. [8] [10]

It was known from the X-ray crystallography that the heavy atoms C, N, and O of each urea molecule lie in the same plane with $C_{2v}$ symmetry, analogous to water.[11] The planar structure of urea in solid state is further confirmed by polarized infrared spectroscopy. [12] The previous assignment of phase V urea to a centrosymmetric space group $Pnma$ is intriguing and rather counterintuitive for an acentric urea molecule. Nonetheless, the phase transition of ice VII into centrosymmetric ice X with $Pn3m$ space group has been theoretically predicted and observed in several spectroscopy studies.[13][14][15] It is proposed that symmetric ice phase X forms at a sufficiently high pressure (up to 62.1 GPa) as the hydrogen bonded protons relocate to the symmetric midpoints of the O-O seperations. [13][16][17] Moreover, a pressure induced order-disorder phase transition was observed for the ices VII and VIII, prior to the symmetrization. [18] This first order phase transition (driven by proton tunneling) is not a true thermodynamic phase change but rather a more gradual phenomenon describing the crossover from a molecular solid to a covalent/ionic solid where short range proton order has disappeared.[18] In fact, the symmetrization of hydrogen bond, which is directly related to the quantum tunneling of protons along the hydrogen bonded axis, has been suggested to exist in hydro-nitrogen solids.[19] Recent discovery reveals an analogous quantum symmetrization occurs in $H_3S$ system and form a sulfur hydride superconductor phase ($Im\bar{3}m$) with high $T_c$ of 203 K. [20] Similar to water, in the H-bonded urea crystals, another pressure-sensitive order-disorder phase transformation associated with mobile proton transfer is expected to exist.[21] For example, a rich variety of short, strong hydrogen bonds (r(O-O) = 2.4 $\mathring{A}$) with high proton mobility was found in urea-acid complexes. [21] Neutron studies have clearly demonstrated migratory behavior of the bridging proton (H atom) toward the center of the bond with temperature increase.[22] [23] If the observed pressure-induced phase transformation into a centrosymmetric structure above 7.2 GPa is accurate, urea may provides a unique opportunity for the study of the dynamics of the 2D proton lattice coupled with charge transfer under a much lower pressure.[6] Experimental conditions under moderate pressure range (within 20 GPa) may enable spectroscopy measurements that weren't able to performed on water due to the small sample sizes, such as the far-infrared spectroscopy. Recent Mid-IR spectroscopy study also reveals the unique pressure-induced softening of the $NH_2$ symmetric bending vibrations between 3 and 8 GPa [24], and an abrupt blue shifts of the frequency at above 8 GPa. This is attributed to be relevant to the dramatic change of hydrogen bonds environment associated with the transformation into phase V.[24] Nonetheless, without crystal structure information for the positions of the hydrogen atoms, the origins of these phenomena had to be inferred. [24]

In contrast to several high-pressure studies at room and/or high temperature, the structure and spectroscopic characteristic of the centrosymmetric phase V has not yet been unambiguously confirmed.[24] [6] [5] In this study, we have performed second harmonic generation (SHG) measurements, powder X-ray diffraction (PXRD), Raman spectroscopy, and an extensive infrared absorption spectroscopic study (both mid-infrared and far-infrared) aimed to explore the high-pressure phases of urea, especially phase V. We have used Universal Structure Predictor: Evolutionary Xtallography (USPEX) based on evolutionary approach to get insight into the polymorphism of urea and predicted the centrosymmetric phase(s) V of urea under high pressure along with metastable polymorphs. Diamond anvil cell uniquely provides optical access to the sample from X-ray to far-IR, which enables in-situ measurements on the exotic centrosymmetric phase. Through these measurements, we have confirmed that the pressure-induced transformation at room temperature is kinetically controlled and a centrosymmetric polymorph dominates above 10 GPa. It is also observed that the charge transfer interaction is very strong at the range of 5.2 to 10 GPa, which could be associated with a disorder complex originate from proton tunneling. Through a combined theoretical and experimental investigation, a new phase diagram for room-temperature compression have been obtained up to 20 GPa.

# 2 Method

## 2.1 Sample Loading

Crystalline urea was obtained from Sigma Aldrich in high purity (99%) and was loaded into DACs with culet sizes ranging between 300 and 500 microns. The DAC was prepared by indenting a tungten gasket to a thickness of 50 to 80 microns. The sample chamber was prepared by drilling a hole into the center of diamond indentation. We used the reference scale by Mao et al. for the determination of pressures in non-hydrostatic pressure conditions.[25]The sample was loaded without pressure medium to avoid forming inclusion compounds except for in the infrared spectroscopy measurements.

## 2.2 Second Harmonic Generation

Second harmonic generation (SHG) measurements were performed on a powder sample of urea at different pressures. The powder SHG method was developed by Kurtz and Perry.[26]This method is commonly used to estimate the nonlinear optical properties of new materials, to detect the absence of an inversion center in crystalline structures or to detect ambiguous structural phase transitions between the centrosymmetric and non-centrosymmetric space groups.[27] In our experiments, a Q-switched Nd:YAG laser (1064 nm, 5-6 ns, 2 kHz) was used for the generation of the fundamental pump wave. With a harmonic separator, a short-pass filter, and with an interference filter, the fundamental infrared light was separated from the generated second harmonic (532 nm). The generated SHG signal was collected with a photo-multiplier and oscilloscope. The measured intensities were corrected by background signals which are collected between the laser pulses. On each position 640 pulses were measured and averaged. In order to check its homogeneity of the sample, the SHG signal was collected from three different areas at most pressure points.

## 2.3 X-ray Scattering

X-ray diffraction patterns were collected at P02.2 Extreme Conditions Beamline (ECB) at PETRA III.[28] A monochromatic beam with the wavelength of 0.4828 $\AA$ was focused on the sample, and diffraction pattern were recorded using 2D flat panel detectors from Perkin Elmer (XRD1621) with an exposure time of 5 or 10 seconds. Prior to experiments, a pressure cycle is performed on the urea sample in order to obtain good powder averaging statistics. The effect of pressure-induced re-crystallization, which gives rise to a good powder sample, can be found in supplementary information. 2-D diffraction patterns were integrated into 1-D patterns with intensity versus 2Theta using the data processing software DIOPTAS. [29] The high energy X-ray beam was focused to the size of 2 x 2 micron$^2$ using Kirkpatrick-Baez mirrors.

### 2.4 Infrared and Raman Spectroscopy

A symmetric DAC including an ultra-low fluorescence synthetic type IIa diamond anvils with 500 microns culet is employed for simultaneous measurements of high-pressure infrared and Raman spectra. Both the Raman and infrared spectroscopy is performed at the Frontier Synchrotron Infrared Spectroscopy (FIS, 22-IR-1) beamline at the National Synchrotron Light Source II (NSLS-II) at Brookhaven National Laboratory. A T-301 stainless steel gasket was preindented to 40 microns thick and a 200 microns hole was drilled in the center of the indentation. The size of the sample chamber is selected based on the diffraction limit of the longest wavelength far-IR light used (50 cm$^{-1}$) in the measurements. The urea sample powder is squeezed into a thin pellet (less than 10 microns) using the DAC in order to avoid saturation of absorption bands. The sample pellet is then loaded into the sample chamber with KBr as a pressure medium for mid-IR measurements. Infrared spectra are collected using a Bruker Vertex 80V FTIR spectrometer and a Hyperion 2000 IR microscope with a liquid nitrogen cooled MCT detector. For far-IR measurements, Vaseline is used as a pressure transmitting medium and a liquid-helium-cooled bolometer is used for signal detection.

### 2.5 Crystal Structure Prediction

First principles crystal structure prediction calculations have been carried out using Universal Structure Predictor: Evolutionary Xtallography (USPEX)[30–32] code based on constrained evolutionary approach.[] We performed an extensive molecular crystal structure search for Urea at distinct pressures such as $\sim$ 0, 5, 10, 15 and 20 GPa with 2, 4 and 6 formula units per unit cell. The first generation with 100 structures are randomly generated and the succeeding 44 generations with a population size of 50 were obtained by applying heredity (30%), random (60%), and rotational mutation (20%) operators until the best structure remains invariant up to 15 generations. The first principles calculations were performed within the framework of density functional theory (DFT). To obtain the global minimum energy structures, the structural optimization with projector-augmented plane-wave (PAW) potentials[33], within the generalized gradient approximation of Perdew-Burke-Ernzerhof (PBE) parametrization[34] as implemented in the Vienna Ab-initio Simulation Package (VASP).[35] A kinetic energy cutoff of 560 eV was used for the plane wave basis set expansion and also $2\pi \times 0.024$ $\mathring{A}^{-1}$ k-spacing has been chosen to sample the Brillouin zone.

# 3 Results

## 3.1 Second harmonic generation

In order to clarify the phase evolution of urea under pressure and resolve the dispute in literature[24] [7] [6] [5], we carried out second harmonic generation (SHG) experiments. SHG, as a nonlinear optical process, is only allowed in crystalline structures with no inversion center symmetry. The application of SHG measurements in high-pressure research has been recently summarized by Bayarjargal and Winkler.[36] Furthermore, it has proven to be extremely useful for studying the pressure-induced structural changes in hydrogen-bonding molecular solids such as ice phase VII.[36] Molecular urea molecule exhibit large electric dipole moment in solution (4.56 D).[37] It has also been demonstrated that the non-centrosymmetric tetragonal phase I urea ($P42_1m$) is an excellent nonlinear optical crystal for SHG, with nonlinear coefficient substantially greater than crystalline quartz.[26] Previously, the efficiency of SHG in urea was measured to 8.5 GPa, using a diamond anvil cell.[38] A sharp peak in the efficiency was found at 0.5 GPa, which clearly indicates a phase transition. However, there was no evidence for centrosymmetric phases that would be apparent as the SHG signal vanish.[38] This is inconsistent with the previous assignment of *Pnma* symmetry to urea phase V above 7.2 GPa by Weber et al. [6] As a result, we re-investigate the non-linear optical properties of urea in order to elucidate the symmetry of its high-pressure phases.

The SHG intensity was collected in two subsequent compression and decompression cycles between 0 and 15 GPa. In the initial compression run, the pressure-induced phase transitions were observed at increasing pressure around 0.5 GPa, 5.3 GPa, and 10 GPa as shown in Figure S1. After the first compression cycle, the urea sample was re-compressed and the sample was observed to exhibit good powder statistics after the pressure treatment. In Figure 1, pressure induced phase transitions were observed at increasing pressures around 0.8 GPa, 5.2 GPa, and 10 GPa.The re-compression data is presumed to be representative of the thermodynamic behavior of urea molecules under pressure as we expect the kinetic meta-stability effect will be minimized when the sample is composed of fine grain size crystals with no preferred orientation. This can be understood by the fact that a reconstructive phase transition typically reach equilibrium faster when there's less diffusion barriers. Observations from the X-ray diffraction measurements which show that the re-compressed sample has a much better powder statistics is shown in Figure S2. Previously, it was observed that SHG intensity slightly increase by a factor of 1.2-1.65 from ambient to 0.3 GPa. Then, a precipitous drop occured upon further compression. At 4.5 GPa, a broad maximum was observed in an urea crystal sample.[38] The first-order phase transition in urea at 0.5 GPa is well established based on p-V-T or $C_p$ measurements, which is also consistent to the sharp drop in SHG signal near the pressure range observed in this work. Nonetheless, we do not observe any indication of the "broad maximum" at 4.5 GPa, but rather an abrupt slope change at 5.2 GPa.[38] It's noted that the broad maximum previously observed at 4.5 GPa appears to be only evident in crystals but rather ambiguous in powder sample. As a result, the overall features are similar compared to literature although the observed phase transition pressure 5.2 GPa is higher than the reported 4.5 GPa. This could be possibly attributed to the kinetic stability of urea phase III as the 4.5 GPa was determined based on the average of six runs, which is also evidenced by the hysteresis observed in decompression. In the present work, based on the two compression runs, the average phase transition pressure are determined to be at 0.8 GPa, 5.2 GPa, and 10.1 GPa.

Intriguingly, it is found that the phase transitions at 0.8 GPa, 5.2 GPa are reversible phase changes, whereas the phase transition at 10 GPa is irreversible and show a significant hysteresis of 3GPa. The ambient phase ($P\bar{4}2_1m$, Z = 2) transforms into orthorhombic phase III ($P2_12_12_1$, Z = 4) at 0.5 GPa, consistent with Raman studies by Lamelas et al. [7] The SHG results are also in excellent agreement with the X-ray observations presented in Figure 2, which will be discussed in a later section.

## 3.2 Polymorphism in urea under high pressure

Urea crystallizes in the tetragonal $P\bar{4}2_1m$ space group with two molecules per unit cell (Z=2) at ambient pressure. This $P\bar{4}2_1m$ phase is commonly referred to as urea phase I. Previous neutron study suggested that the crystal structure of high-pressure phases of urea can be indexed to (1) Phase III ($P2_12_12_1$; Z = 4) from 0.5 to 2.8 GPa), (2) Phase IV ($P2_12_12$; Z = 2) from 2.8 to 7.2 GPa, and (3)Phase V (Pnma; Z = 4) from 7.2 to 9.0 GPa and beyond. With increasing pressure, it was observed that each new phase is a progressively more distorted version of phase I.[6] We note that the crystal structure of phase II was sometimes referred to as phase IV but they are in fact identical structure.[3] [24] [7]Although most researchers in the field follow this proposed sequence, in this work, we have a similar observation of a sequence of four structural phases occurring over the pressure range from 0 to 15GPa at room temperature as Lamelas et al.[7] Phase I ($P\bar{4}2_1m$) is found to phase transform into phase III ($P2_12_12_1$) at 0.5 GPa if the sample was held under pressure for 20 minutes (see Figure S3). In addition, lattice parameters and space group of two additional high-pressure phases are determined *ab initio* using DICVOL.[39] For example, at 8 GPa, phase X is identified and tentatively indexed to a orthorhombic lattice a=4.700, b=7.619, c=9.010 with space group $Pnnm$. We note that the index is not unambiguously confirmed due to its disorder nature. Most interestingly, the centrosymmetric phase $V'$ is indexed to a slightly distorted monoclinic lattice with lattice parameters a=4.692, b=7.206, c=8.885, and beta=88.76 with the space group $P2_1$/m at 11.9 GPa, in excellent agreement with the SHG results which requires a centrosymmetric space group . As shown in Figure 2, the Pawley method is used for full-profile refinement of the PXRD pattern obtained at 11.9 GPa.[40] The refinement was performed in GSAS-II, which gives lattice parameters provided above and extracting intensities.[41] The inset image plate shows the 2D diffraction pattern with complete powder averaging statistics, which offer reliable extracted peak intensities values which can then be used in the structure solution program Endeavour.[42] The weighted residual wR value is calculated to be 15.6 % and the unweighted phase residuals $RF^2$ is 2.3 % on 58 reflections. This refinement result is convincing accounting for the strain-induced peak broadening under pressure. Two structure solutions for phase $V'$ urea with $P2_1$/m symmetry independently derived from theory and structure annealing program are proposed in this work for the first time (See Figure S4 and Supplementary Information for structural models).

As shown in Figure 3, the transition from phase III ($P2_12_12_1$) to a orthorhombic phase X ($Pnnm$) at 6.5 GPa appears to be a progressively distorted version of phase III. Similar to the SHG results, it is observed that the back transformation at phase $V'$ occurs around 6.9 GPa. Compared to the transition pressure at 9.9 GPa during compression, there's a similar 3.3 GPa of hysteresis. As a result, the phase change between X/$V'$ is considered a first-order phase transition with expected meta-stability of the centrosymmetric phase $V'$.[43] This phase change could be a displacive rather than a reconstructive phase transition, which could also show hysteresis.[43] As shown in Figure S5, the hydrogen bond lattice vibrations transition from X to $V'$ rather smoothly with the major difference being the intensity change of the lattice mode $E_2$. Compared to the spectral characteristic which show total reconstruction from phase I to phase III (see Figure S6), it is clear that the change of features from X to $V'$ is uniquely continuous and mild. Given the similarities of the XRD patterns and lattice phonon behaviors in Figure S5, it is an interesting possibility to consider that this phase change between X/$V'$ could be a first-order and displacive phase transition, if phase X is not disorder. This is further supported by the trend observed in the unit volume per urea molecule as a function of pressure and the fitted equation of state of phase III and $V'$ presented in Figure 5. Nonetheless, without the crystal structure solution of phase X, the nature of this phase transformation is inconclusive.

The equation of state of high pressure phases of urea also provides critical information of the structural evolution. As shown in Figure 5, the theoretical predicted equation of state of phase I, III, $V'$ are plotted in dash lines and compared to the experimental observations indicated by circle markers. Phase I, III, $V'$ are plotted in black, blue, and red color respectively. The unit volume of phase X is marked by empty red circles to represent that it has a similar but less distorted orthorhombic lattice of phase $V'$. It is clear that there's an abrupt volume drop between phase I and III, which supports the fact that it's a first-order phase change as reported previously.[5] This can also be directly observed from the drastic change of PXRD patterns as shown in Figure S6, which suggests a total reconstruction of H-bonding networks. Nonetheless, the volume change at the transition between phase III and X as well as phase X and $V'$ appears to be relatively small as suggested by both experiment and theory. This is evidenced by the continuous change of diffraction patterns with increasing pressure as shown in Figure 3 starting from 6.5 GPa. The hysteresis observed in the SHG measurements has ruled out the possibility of having a secondary phase change from phase X to $V'$.[43] However, the volume change between these high-pressure phases appear to be very small, which is also observed in theoretical results.

### 3.3 Vibrational spectroscopy and lattice phonon dynamics

Figure 6 is a plot of Raman and IR frequencies of the lattice (intermolecular) modes from ambient pressure to approximately 15 GPa. In Figure 6 (a), typical Raman and IR spectra obtained from urea powder sample at 10 GPa are presented. The two distinct spectra collected at 10 GPa unambiguously shows the mutually exclusive Raman and IR-active modes, which further confirm the centrosymmetric crystal lattice symmetry of urea phase $V'$. As shown in Figure 6, the mutually exclusive selection rules of Raman and IR-active vibration do not show up until 10 GPa, at the same pressure range where SHG signal disappear. The Raman and IR frequencies of phase I, III, and X lattice modes are also plotted in Figure 6 with the phase regions marked by red, blue, and purple respectively following the color code in Figure 1.

Urea was chosen as the subject of multiple vibrational spectroscopy studies as a model hydrogen-bonded system. In particular, the evolution of lattice modes under pressure may provide critical insights to the high-pressure phases of urea. Previously, limited infrared[44] and Raman[45] studies were performed to study the I-III phase transition at 0.5 GPa. Lamelas et al also performed an extensive Raman experiments which included spatically resolved measurements that were used to generate maps of the Raman spectra collected from single-crystal and powder sample up to 12 GPa. [7] Nonetheless, it was often observed two coexisting spatial domains that give two distinct spectra at a given pressure starting from 5.0 GPa and beyond. This inhomogeneous vibrational characteristic was surprisingly observed even in single-crystal sample, regardless of which pressure medium (Ar, $N_2$, and mineral oil) was used.[7] As a result, the high-pressure phase diagram becomes ambiguous due to the local probe of Raman spectroscopy tool, which makes it difficult to determine the dominating phase at a target pressure. Here, we propose to use combined Raman and far-infrared absoprtion spectroscopy with synchrotron light source to study the high-pressure phase evolution. With the supplement of information from the far-IR, it not only shed fresh insights on the selection rules of the lattice vibration modes but also provide bulk analysis as the far-infrared beam is approximately 200 microns in size in order to cover the spectral range from 50 to 400 $cm^{-1}$, which covers the entire diamond cell chamber. Furthermore, far-IR is also generally a very useful tool to study hydrogen bond vibrations and charge transfer. For example, hydrogen bonded phenol complexes show intermolecular lattice vibrations in the range of 50 $cm^{-1}$ to 200 $cm^{-1}$, but such external vibrations were not observed in the case of crystalline aromatic hydrocarbon complexes such as pyrene.[46] In contrast to Raman, the absolute intensities of these intermolecular vibrations in the far-IR spectra are directly relevant to the transition moment function. As a result, the intensity change of the far-IR bands associated with hydrogen bond vibrations could indicate a H-bonding network formation or reorganization. For instance, Pietrzycki *et al* recorded and calculated the IR spectrum of quinhydrone complex and discovered that the shifts of the absorption bands after the formation of the complex could originate from the charge transfer interaction of hydrogen bonds. [47] The high-pressure Raman and IR measurements were performed simultaneously at the Frontier Synchrotron Infrared Spectroscopy beamline (22-IR-1) located at the National Synchrotron Light Source-II, with the end stations uniquely optimized for the study of materials at extreme pressures.

A summary of all Raman and IR-active modes in phase I is presented in Table S1. At ambient pressure, the lattice-mode observed in this work is in excellent agreement with earlier experimental data.[48] The measured frequencies are also described well by a previously proposed model utilizing a phenomenological Buckingham-type potential.[49] Among the five Raman-active lattice modes, three modes including $E_1$, $E_2$, and $E_3$ located at 104 $cm^{-1}$, 135$cm^{-1}$, and 187 $cm^{-1}$ are also IR-active at ambient. The selection rules show a drastic change at the phase transition at 0.5 GPa. For example, the Raman-active only $B_1$ mode at 59 $cm^{-1}$ becomes IR-active only in the phase III pressure range (0.8 GPa to 6.5 GPa). On the other hand, the $E_3$ mode, which is both Raman and IR-active, becomes Raman inactive in the same pressure range but become both Raman and IR active again in the pressure range of 8 to 10 GPa. This is consistent with the presumption that the disorder phase X which appears in the pressure range of 5.2 GPa to 8 GPa, but there might be a two-phase regions from 8 to 10 GPa, in which the phase X is not dominating anymore. As a result, the $B_1$ mode is not captured in the Far-IR spectra beyond 8 GPa as it might just be a minority phase in the powder bulk sample probed. However, the $E_2$ mode remains IR-active and their IR activity holds all the way through compression up to 15 GPa. As a result, its intense FIR absorption band is selected to be a marker to monitor the high-pressure phases behavior in the bulk sample. As shown in the time-dependent lattice vibration analysis (See Figure S3), the FWHM of the far-IR absorption band associated with $E_2$ lattice phonon varies with pressure increase. An anamolous increase of FWHM appears to occur around 0.6-0.8 GPa and 8 GPa at the I-III and X-$V'$ phase transitions. This indicates that there might be a mixed two-phase regions at these pressure range, consistent with the SHG results. Furthermore, it is also noticed that the absolute intensities of the $E_2$ mode are continuously changing when the sample is held at a target pressure in the pressure range of 6 to 10 GPa. As shown in Figure S3, the time-dependent infrared spectra obtained at 7.4 GPa show saturated $E_2$ absorption band when the target pressure is first reached. Subsequently, the intensity decreases with time and saturation is not observed after the pressure is held at 7.4 GPa for 15 minutes. As a result, the peak width of $E_2$ is obtained at this stage. Nonetheless, we further observed a saturation when the sample is held under pressure for 20 minutes. This suggests that the H-bonding system in the pressure range of 6 to 10 GPa is dynamic and system is not able to reach equilibrium within a short time frame. We have also attempted to let the sample sit under pressure overnight. However, no sign of reaching equilibrium is observed. Therefore, this kinetic study provides excellent support for the conclusion that phase X is a disorder phase. Previously, Dziubek *et al* observed a softening red shift of the $NH_2$ symmetric rocking mode between 3 and 8 GPa under room-temperature compression.[24] In this work, we also see similar behavior in the MIR range (see Figure S5). We are leaning toward attributing this to a charge transfer phenomenon, which is well-understood in an analogous quinhydrone crystal system.[46][47][10] The $NH_2$ symmetric rocking mode shifts to a lower energy until it reaches an abrupt slope sign inversion point at 8 GPa, which is the onset of formation of phase V. This suggests that the phase III urea lattice becomes more ionic. [10] The spectral change in the $NH_2$ rocking mode clearly has parallels to the drastic intensity changes of the lattice $E_2$ mode observed in the FIR range. Similarly, the intensity of the saturated $E_2$ absorption band starts to drop beyond 8.4 GPa. Origins of the charge transfer phenomenon could include both electron transfer and proton transfer, as discussed in details in the next section.

# 4 Discussions

In this work, we do not observe any evidence of transition from III to IV at 2.8 GPa or IV to V at 7.2 GPa as indicated by previous neutron observations. [6] Therefore, we propose that the widely accepted phase transition sequence I ($P\bar{4}2_1m$, Z = 2), III ($P2_12_12_1$, Z = 4), IV ($P2_12_12$, Z = 2), and V(Pnma, Z = 4) at 0.5 GPa, 2.8 GPa, and 7.2 GPa needs to be reconsidered.[6] [24] Instead, based on the single-crystal Raman study [7] and SHG results, we propose that the transitions to the high pressure phases on average take place at pressure near 0.8 GPa, 5.2 GPa, and 10.1 GPa.[7] It is suspected that the use of gas phase pressure medium might be the reason for the inconsistent observations of high-pressure phase behaviors in the literature.[24][5] For example, Strobel *et al* has proposed that urea could be a potential hydrogen storage material. [50] Although it has been shown that urea does not form clathrates with deuterium within 3.7 GPa by high-pressure neutron diffraction, it is possible that urea can form guest-host clathrate compounds under pressure with commonly used inert gas pressure medium such as Ne. [2]

Moreover, the absence of SHG signals beyond 10 GPa affirmatively shows a phase transformation of a non-centrosymmetric structure to a centrosymmetric structure. The intensity of the SHG is the transmitted light of second harmonic wave ($\lambda_{2\omega}$ = 527 nm) of a finely ground urea powder when excited with an incident pulsed laser fundamental wave ($\lambda$=1054 nm). The detection of second harmonic photon that is due to the conversion in the bulk sample unambiguously excludes centrosymmetry. As a result, as long as solid urea remains in the non-centrosymmetric orthorhombic space group, the SHG signal is presented. However, the SHG signal dramatically decreases starting at  5.2 GPa and disappears at 10.1 GPa, which suggests a phase transition from a noncentrosymmetric space group to a centrosymmetric space group. Previous neutron studies reported that urea crystallizes into Phase V (*Pnma*; Z=4) at the pressure range of 7.2 GPa to 9.0 GPa. [6] *Pnma* is a centrosymmetric space group and therefore we follow the convention and assign the centrosymmetric phase beyond 10.1 GPa as phase V. However, in contrast to previous report, here centrosymmetry is unambiguously excluded at the pressure range of 7.2 to 9.0 GPa in the experiment (pink region).[6][24] This could be explained by the hypothesis that it is a disorder phase region where the crystal structure and symmetry continuously change with increasing pressure. In addition, we also observed that the back phase transition of the phase V during decompression is at 6.9 GPa, indicating a hysterisis of 3.2 GPa. If we take the average of the observed phase transition pressure in the compression (10.1 GPa) and the decompression (6.9 GPa), the average phase transition pressure at equilibrium is about 8.5 GPa. In the case of a first-order and reconstructive phase transition, which is common for hydrogen-bonding motif, we expect meta-stability of the phase V at decreasing pressure. Therefore, it is sensible to consider the pressure region from 8.5 to 10.1 GPa as a two-phase region, which would include a mixture of phase V and a lower pressure disorder phase. In this case, it is also possible that previous work might have also mistakenly indexed a powder diffraction pattern that contains two phases into a single unit cell.[6] [24] In particular, reported indexing of urea crystal structure beyond 10 GPa is rare.[7] High-pressure crystallography based on a limited pressure range could give an incomplete and often misleading picture of the phase diagram of urea.

To solve the puzzle, we have performed an extensive theoretical crystal structure search to explore the polymorphism in urea using USPEX package at different pressures ∼0, 5, 10, 15 and 20 GPa with 2, 4 and 6 f.u. per primitive cell. Our CSP calculations reproduce the low pressure well known polymorphs of urea namely phase I, III and IV and also predicted new stable (V and V′) and meta stable (III′) polymorphs for urea which are not known before. As illustrated in figure 4, phase I undergoes a first order structural phase transition to phase III at ∼ 0.32 GPa, the transition pressure is underestimated when compared to the experimental transition pressure of 0.48 GPa[24] and theoretical transition pressure of 0.66 GPa[51]. We also predicted a new polymorph III′ which is proposed based on previous X-ray diffraction and Raman spectroscopic measurements, however, this phase is found to be meta stable in the studied pressure range. Upon further compression, the phase III expected to undergo to phase IV based on the previous high pressure experimental and theoretical studies on Urea. Interestingly, the phase IV is found to be metastable throughout the studied pressure range (0-20 GPa). Based on HP Raman and XRD measurements, Bridgman[] and Dziubek *et al*[24] proposed that the phase II and IV are in fact the same and phase II/IV is observed at above room temperatures (RT) in their studies which clearly demonstrates that phase II/IV is a high temperature polymorph of urea. In the present study, we didn't observe phase IV neither with RT HPXRD nor spectroscopic measurements and these results are consistent with our present theoretical calculations. Moreover, the predicted phases such as phase V and V′ at T = 0 K are energetically competitive over the pressure range of 10-20 GPa. Although, the phase V′ seems to be stable at T = 0 K with ∼ 2.5 meV/atom over phase V. The kinetic effects plays a significant role in determining the stability of stable polymorph of urea, as we observed phase V in the RT HPXRD measurements. Thus, the predicted crystal structures provide an insight for the present and future experimental and theoretical studies on urea polymorphs.

An alternative possible index considered is that phase X could be a mixed phase of phase III ($P2_12_12_1$) with lattice parameters of a=3.317, b=8.138, and c=8.743 and phase IV ($P2_12_12$) with a=3.433, b=7.318, c=4.573. The peak positions appear to be close, however, the possibility of a mixed phase of III and IV is ruled out based on the unique infrared characteristic of disorder hydrogen bonded solid observed in the pressure range of 6.5 to 10 GPa. Details were discussed in the vibration spectroscopy and phonon dynamics section.

As the SHG measurements clearly show, there is a possible co-existence of non centrosymmetric and centrosymmetic phases of urea in the pressure range of ∼ 6-10 GPa. From the present theoretical and experimental studies, we propose two possible non centrosymmetric and centrosymmetic phases in this pressure range *i.e.*, 1) $P21$-$P2_1/m$ and 2) $Pna2_1$-$Pnma$. As pressure increases, the non centrosymmetric structure gradually transforms to one of the above proposed centrosymmetic structures.

Regarding the disorder observed in phase X, an interesting hypothesis of the positively charged proton transfer is discussed below. Analogous to the hydroquinone crystal, the intermolecular proton transfer takes place accompanied by a rearrangement of the electron configuration of the urea molecules into its ionic state.[10] [52] Two competing interactions are involved in this process; (1) the net dissociation energy of protons from $NH_2$ to $N^{2-}$, $E_d$ , and (2) the electrostatic Madelung energy of the

ionized lattice, M. At ambient pressure, the potential well for proton (H) is considered to be asymmetric because of the large energy difference ($E_d$-M), while the proton tunneling between the bottom of the well and the neighbor local minimum is small. To be more specific, the hydrogen atom is involved in two partial covalent bonds (i.e., N-H-O); one convenient description of the potential energy profile defining this type of H-bond is as a broad, asymmetric quantum well.[53] However, the potential well becomes more symmetric upon the application of pressure, which results in a higher probability of proton tunneling as ($E_d$-M) approaches zero.[10] As a result, Phase X observed in the range of 6 to 8 GPa is likely a disorder phase result from the quantum melting of the hydrogen-bonding lattice because of its tunneling motion. Previously, it was observed in urea phosphoric acid co-crystal that a very short hydrogen bond via an O-O linkage (2.417 $\AA$) displays high proton mobility, which leads to the proton position change by 0.05 $\AA$ toward the center of the O-O axis as temperature increase from 15 to 300K. [21] While the N-O linkage of urea phase V is still linked through standard hydrogen bonds (rN-O= 2.779 $\AA$) and no apparent bond symmetrization was observed according to our DFT calculation performed at 0K. It is possible that the proton mobility will be significantly enhanced at 300K. Nonetheless, without the hydrogen positions obtained from neutron diffraction, this charge transfer process has to be inferred.

## 5 Conclusions

We employ high-pressure second-harmonic generation (SHG) to establish unambiguously that the high-pressure polymorph $V'$ of acentric urea adopts a centrosymmetric structure, analogous to the symmetric hydrogen-bonding network in ice phase X. Combined with synchrotron far-infrared spectroscopy, the SHG measurements reveal signatures of charge transfer in the pressure range of 6–10 GPa, suggesting disordering from the quantum melting of the hydrogen-bonding lattice. These results indicate that noncentrosymmetric and centrosymmetric structural motifs may coexist in this pressure range, potentially accounting for the discrepancies reported in early high-pressure powder-diffraction studies. Above 10 GPa, the lattice-phonon selection rules become mutually exclusive, coincident with the disappearance of the SHG response. This concurrence provides strong evidence for a transition to a centrosymmetric phase. By combining SHG, synchrotron far-infrared spectroscopy, and crystal-structure searches using the Universal Structure Predictor: Evolutionary Xtallography (USPEX), we construct a revised high-pressure phase diagram of molecular urea at room temperature.

==========================================

**Acknowledgements.** This work was supported by the Office of Research and Development (ORD) at National Taiwan University, the National Science and Technology Council (NSTC) under Grant No. NSTC 114-2221-E-002-149, and the Joint Photon Science Institute at Stony Brook University. The high pressure diffraction experiments were carried out at the Extreme Conditions Beamline P02.2 at PETRA III at DESY in Hamburg, Germany. DESY is a member of the Helmholtz Association HGF. The high pressure infrared spectroscopy measurements were conducted at the Frontier Synchrotron Infrared Spectroscopy beamline (22-IR-1, FIS) at the National Synchrotron Light Source II, a U.S. Department of Energy (DOE) Office of Science User Facility operated for the DOE Office of Science by Brookhaven National Laboratory under Contract No. DE-SC0012704. The authors acknowledge Dylan Darnulc's help in collecting the XRD data as well as the Joint Photon Science Institute at Stony Brook University for travel support.

## References

[1] Hopf, H.: The life and work of friedrich wohler (1800-1882). von robin keen. Angewandte Chemie **117**, 0044–8249 (2005)

[2] Donnelly, M., Bull, C.L., Husband, R.J., Frantzana, A.D., Klotz, S., Loveday, J.S.: Urea and deuterium mixtures at high pressures. The Journal of Chemical Physics **142**(12), 124503 (2015) https://doi.org/10.1063/1.4915523 https://doi.org/10.1063/1.4915523

[3] Olejniczak, A., Ostrowska, K., Katrusiak, A.: H-bond breaking in high-pressure urea. The Journal of Physical Chemistry C **113**(35), 15761–15767 (2009) https://doi.org/10.1021/jp904942c https://doi.org/10.1021/jp904942c

[4] Bridgman, P.W.: Polymorphism at high pressures. Proc. Amer.Acad. Arts Sci. **52**, 91–187 (1916)

[5] Moses Abraham, B., Adivaiah, B., Vaitheeswaran, G.: Microscopic origin of pressure-induced phase-transitions in urea: a detailed investigation through first principles calculations. Phys. Chem. Chem. Phys. **21**, 884–900 (2019) https://doi.org/10.1039/C8CP04827D

[6] Weber, H.P., Marshall, W.G., Dmitriev, V.: High-pressure polymorphism in deuterated urea. (2002)

[7] Lamelas, F.J., Dreger, Z.A., Gupta, Y.M.: Raman and x-ray scattering studies of high-pressure phases of urea. The Journal of Physical Chemistry B **109**(16), 8206–8215 (2005) https://doi.org/10.1021/jp040760m https://doi.org/10.1021/jp040760m. PMID: 16851959

[8] Horita, J., Santos, A.M., Tulk, C.A., Chakoumakos, B.C., Polyakov, V.B.: High-pressure neutron diffraction study on h–d isotope effects in brucite. Physics and Chemistry of Minerals **37**(10), 741–749 (2010) https://doi.org/10.1007/s00269-010-0372-5 https://doi.org/10.1007/s00269-010-0372-5

[9] Catti, M., Ferraris, G., Hull, S., Pavese, A.: Static compression and h disorder in brucite, mg(oh)2, to 11 gpa: a powder neutron diffraction study. Physics and Chemistry of Minerals **22**(3), 200–206 (1995) https://doi.org/10.1007/BF00202300 https://doi.org/10.1007/BF00202300

[10] Mitani, T., Saito, G., Urayama, H.: Cooperative phenomena associated with electron and proton transfer in quinhydrone charge-transfer crystal. Phys. Rev. Lett. **60**, 2299–2302 (1988) https://doi.org/10.1103/PhysRevLett.60.2299

[11] Vaughan, P., Donohue, J.: The structure of urea. Interatomic distances and resonance in urea and related compounds. Acta Crystallographica **5**(4), 530–535 (1952) https://doi.org/10.1107/S0365110X52001477

[12] Keller, W.E.: Evidence for the planar structure of the urea molecule. The Journal of Chemical Physics **16**(10), 1003–1004 (1948) https://doi.org/10.1063/1.1746669 https://doi.org/10.1063/1.1746669

[13] Aoki, K., Yamawaki, H., Sakashita, M., Fujihisa, H.: Infrared absorption study of the hydrogen-bond symmetrization in ice to 110 gpa. Phys. Rev. B **54**, 15673–15677 (1996) https://doi.org/10.1103/PhysRevB.54.15673

[14] Lee, C., Vanderbilt, D., Laasonen, K., Car, R., Parrinello, M.: Ab initio studies on high pressure phases of ice. Phys. Rev. Lett. **69**, 462–465 (1992) https://doi.org/10.1103/PhysRevLett.69.462

[15] Lee, C., Vanderbilt, D., Laasonen, K., Car, R., Parrinello, M.: Ab initio studies on the structural and dynamical properties of ice. Phys. Rev. B **47**, 4863–4872 (1993) https://doi.org/10.1103/PhysRevB.47.4863

[16] Xu, R., Liu, Z., Ma, Y., Cui, T., Liu, B., Zou, G.: Ab initio investigation of hydrogen bonding and electronic structure of high-pressure phases of ice, arXiv:0801.0400 (2007). https://doi.org/10.48550/arXiv.0801.0400

[17] Goncharov, A.F., Struzhkin, V.V., Somayazulu, M.S., Hemley, R.J., Mao, H.K.: Compression of ice to 210 gigapascals: Infrared evidence for a symmetric hydrogen-bonded phase. Science **273**(5272), 218–220 (1996) https://doi.org/10.1126/science.273.5272.218 https://science.sciencemag.org/content/273/5272/218.full.pdf

[18] Schweizer, K.S., Stillinger, F.H.: High pressure phase transitions and hydrogen-bond symmetry in ice polymorphs. The Journal of Chemical Physics **80**(3), 1230–1240 (1984) https://doi.org/10.1063/1.446800 https://doi.org/10.1063/1.446800

[19] Yedukondalu, N., Vaitheeswaran, G., Anees, P., Valsakumar, M.C.: Phase stability and lattice dynamics of ammonium azide under hydrostatic compression. Phys. Chem. Chem. Phys. **17**, 29210–29225 (2015) https://doi.org/10.1039/C5CP04294A

[20] Errea, I., Calandra, M., Pickard, C.J., Nelson, J.R., Needs, R.J., Li, Y., Liu, H., Zhang, Y., Ma, Y., Mauri, F.: Quantum hydrogen-bond symmetrization in the superconducting hydrogen sulfide system. Nature **532**(7597), 81–84 (2016) https://doi.org/10.1038/nature17175

https://science.sciencemag.org/content/273/5272/218.full.pdf

[21] Morrison, C.A., Siddick, M.M., Camp, P.J., Wilson, C.C.: Toward understanding mobile proton behavior from first principles calculation: the short hydrogen bond in crystalline ureaphosphoric acid. Journal of the American Chemical Society **127**(11), 4042–4048 (2005) https://doi.org/10.1021/ja043327z https://doi.org/10.1021/ja043327z. PMID: 15771541

[22] Wilson, C.C.: Migration of the proton in the strong O—H$\cdots$O hydrogen bond in urea–phosphoric acid (1/1). Acta Crystallographica Section B **57**(3), 435–439 (2001) https://doi.org/10.1107/S0108768100018875

[23] Wilson, C.C., Morrison, C.A.: Structural and theoretical investigations of short hydrogen bonds: neutron diffraction and plane-wave dft calculations of urea–phosphoric acid. Chemical Physics Letters **362**(1), 85–89 (2002) https://doi.org/10.1016/S0009-2614(02)00952-1

[24] Dziubek, K., Citroni, M., Fanetti, S., Cairns, A.B., Bini, R.: High-pressure high-temperature structural properties of urea. The Journal of Physical Chemistry C **121**(4), 2380–2387 (2017) https://doi.org/10.1021/acs.jpcc.6b11059 https://doi.org/10.1021/acs.jpcc.6b11059

[25] Mao, H.K., Xu, J., Bell, P.M.: Calibration of the ruby pressure gauge to 800 kbar under quasi-hydrostatic conditions. Journal of Geophysical Research: Solid Earth **91**(B5), 4673–4676 (1986) https://doi.org/10.1029/JB091iB05p04673 https://agupubs.onlinelibrary.wiley.com/doi/pdf/10.1029/JB091iB05p04673

[26] Kurtz, S.K., Perry, T.T.: A powder technique for the evaluation of nonlinear optical materials. Journal of Applied Physics **39**(8), 3798–3813 (1968) https://doi.org/10.1063/1.1656857 https://doi.org/10.1063/1.1656857

[27] Nalla, V., Medishetty, R., Wang, Y., Bai, Z., Sun, H., Wei, J., Vittal, J.J.: Second harmonic generation from the 'centrosymmetric' crystals. IUCrJ **2**(3), 317–321 (2015) https://doi.org/10.1107/S2052252515002183

[28] Liermann, H.-P., Morgenroth, W., Ehnes, A., Berghäuser, A., Winkler, B., Franz, H., Weckert, E.: The extreme conditions beamline at PETRA III, DESY: Possibilities to conduct time resolved monochromatic diffraction experiments in dynamic and laser heated DAC. Journal of Physics: Conference Series **215**, 012029 (2010) https://doi.org/10.1088/1742-6596/215/1/012029

[29] Prescher, C., Prakapenka, V.B.: Dioptas: a program for reduction of two-dimensional x-ray diffraction data and data exploration. High Pressure Research **35**(3), 223–230 (2015) https://doi.org/10.1080/08957959.2015.1059835 https://doi.org/10.1080/08957959.2015.1059835

[30] Oganov, A.R., Glass, C.W.: Crystal structure prediction using ab initio evolutionary techniques: Principles and applications. The Journal of Chemical Physics **124**(24), 244704 (2006) https://doi.org/10.1063/1.2210932 https://doi.org/10.1063/1.2210932

[31] Lyakhov, A.O., Oganov, A.R., Stokes, H.T., Zhu, Q.: New developments in evolutionary structure prediction algorithm uspex. Computer Physics Communications **184**(4), 1172–1182 (2013) https://doi.org/10.1016/j.cpc.2012.12.009

[32] Oganov, A.R., Lyakhov, A.O., Valle, M.: How evolutionary crystal structure prediction works—and why. Accounts of Chemical Research **44**(3), 227–237 (2011) https://doi.org/10.1021/ar1001318 https://doi.org/10.1021/ar1001318. PMID: 21361336

[33] Kresse, G., Joubert, D.: From ultrasoft pseudopotentials to the projector augmented-wave method. Phys. Rev. B **59**, 1758–1775 (1999) https://doi.org/10.1103/PhysRevB.59.1758

[34] Perdew, J.P., Burke, K., Ernzerhof, M.: Generalized gradient approximation made simple. Phys. Rev. Lett. **77**, 3865–3868 (1996) https://doi.org/10.1103/PhysRevLett.77.3865

[35] Kresse, G., Furthmüller, J.: Efficient iterative schemes for ab initio total-energy calculations using a plane-wave basis set. Phys. Rev. B **54**, 11169–11186 (1996) https://doi.org/10.1103/PhysRevB.54.11169

[36] Bayarjargal, L., B., W.: Second harmonic generation measurements at high pressures on powder samples. Zeitschrift für Kristallographie – Crystalline Materials **229**(2), 92 (2014) https://doi.org/10.1515/zkri-2013-1641 https://www.degruyter.com/view/j/zkri.2014.229.issue-2/zkri-2013-1641/zkri-2013-1641.xml

[37] Kumler, W.D., Fohlen, G.M.: The dipole moment and structure of urea and thiourea1. Journal of the American Chemical Society **64**(8), 1944–1948 (1942) https://doi.org/10.1021/ja01260a054 https://doi.org/10.1021/ja01260a054

[38] Yang, G., Li, Y., Dreger, Z.A., White, J.O., Drickamer, H.G.: High-pressure studies of second harmonic generation in organic materials. Chemical Physics Letters **280**(3), 375–380 (1997) https://doi.org/10.1016/S0009-2614(97)01122-6

[39] Boultif, A., Louër, D.: Powder pattern indexing with the dichotomy method. Journal of Applied Crystallography **37**(5), 724–731 (2004) https://doi.org/10.1107/S0021889804014876

[40] Pawley, G.S.: Unit-cell refinement from powder diffraction scans. Journal of Applied Crystallography **14**(6), 357–361 (1981) https://doi.org/10.1107/S0021889881009618

[41] Toby, B.H., Von Dreele, R.B.: *GSAS-II*: the genesis of a modern open-source all purpose crystallography software package. Journal of Applied Crystallography **46**(2), 544–549 (2013) https://doi.org/10.1107/S0021889813003531

[42] Putz, H., Schön, J.C., Jansen, M.: Combined method for *ab initio* structure solution from powder diffraction data. Journal of Applied Crystallography **32**(5), 864–870 (1999) https://doi.org/10.1107/S0021889899006615

[43] Agarwal, G.S., Shenoy, S.R.: Observability of hysteresis in first-order equilibrium and nonequilibrium phase transitions. Phys. Rev. A **23**, 2719–2723 (1981) https://doi.org/10.1103/PhysRevA.23.2719

[44] M, H.S.L.: Infrared spectra and phase transitions of solids under pressure. i. High Temper.-High Press.; G.B.; DA. 1975; VOL. 7; NO 2; PP. 165-175; BIBL. 39 REF. (1975)

[45] Lascombe, J., Huong, P.V.H., Kielich, S.: Raman spectroscopy: Linear and nonlinear—proceedings of the eighth international conference on raman spectroscopy, bordeaux, france, 6–11 september 1982. wiley hyden 1982, 868 pp., £27.50. Journal of Raman Spectroscopy **14**(3), 219–219 (1983) https://doi.org/10.1002/jrs.1250140319 https://onlinelibrary.wiley.com/doi/pdf/10.1002/jrs.1250140319

[46] Kubinyi, M., Varsányi, G.: Infrared and far infrared spectra of crystalline phenol-p-benzoquinone complexes. Spectroscopy Letters **9**(10), 689–696 (1976) https://doi.org/10.1080/00387017608067459 https://doi.org/10.1080/00387017608067459

[47] Kubinyi, M., Keresztury, G.: Infrared and raman spectroscopic study of molecular interactions in quinhydrone crystals. Spectrochimica Acta Part A: Molecular Spectroscopy **45**(4), 421–429 (1989) https://doi.org/10.1016/0584-8539(89)80037-6

[48] Khilji, M.Y., Sherman, W.F., Wilkinson, G.R.: Variable temperature raman study of urea. Journal of Molecular Structure **143**, 109–112 (1986) https://doi.org/10.1016/0022-2860(86)85216-4 . Proceedings of the XVIIth European Congress on Molecular Spectroscopy,

[49] Bleckmann, P., Thibud, M.: Determination of raman intensities of lattice vibrations in crystalline urea. Journal of Molecular Structure **219**, 7–12 (1990) https://doi.org/10.1016/0022-2860(90)80023-D

[50] Strobel, T.A., Hester, K.C., Koh, C.A., Sum, A.K., Sloan, E.D.: Properties of the clathrates of hydrogen and developments in their applicability for hydrogen storage. Chemical Physics Letters **478**(4), 97–109 (2009) https://doi.org/10.1016/j.cplett.2009.07.030

[51] Moses Abraham, B., Adivaiah, B., Vaitheeswaran, G.: Microscopic origin of

pressure-induced phase-transitions in urea: a detailed investigation through first principles calculations. Phys. Chem. Chem. Phys. **21**, 884–900 (2019) https://doi.org/10.1039/C8CP04827D

[52] Tokura, Y., Okamoto, H., Koda, T., Mitani, T.: Pressure-induced neutral-to-ionic phase transition in ttf-p-chloranil studied by infrared vibrational spectrocopy. Solid State Communications **57**(8), 607–610 (1986) https://doi.org/10.1016/0038-1098(86)90332-7

[53] Koetzle, T.F.: **Single crystal neutron diffraction from molecular materials.** By Chick C. Wilson. Pp. xiii + 370. Singapore: World Scientific, 2000. ISBN 981-02-3776-6. Acta Crystallographica Section A **57**(4), 478 (2001) https://doi.org/10.1107/S0108767301005980

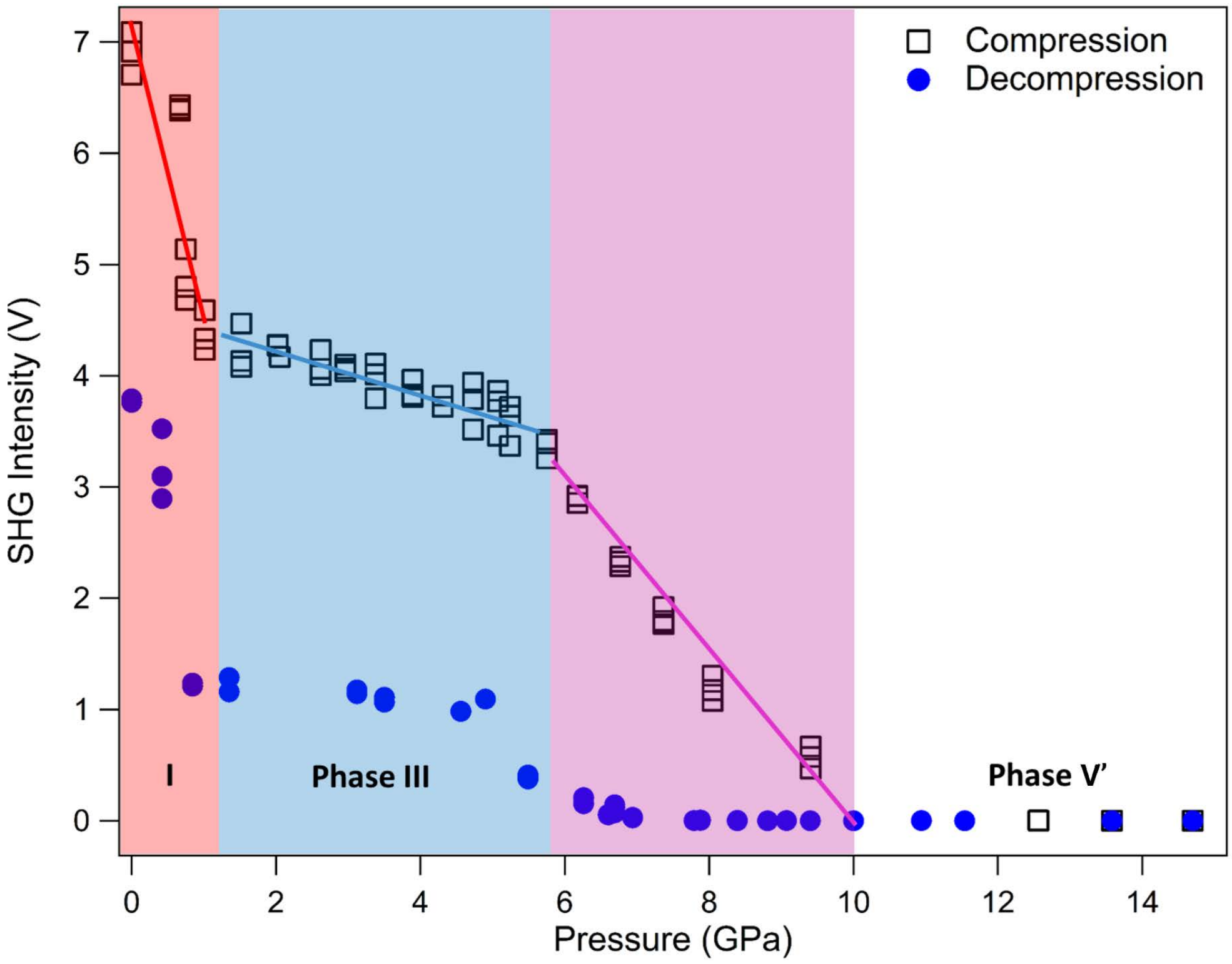

**Fig. 1** Pressure dependence of the second harmonic generation (SHG) signal of urea. Note that the urea sample has been subjected to a pressure cycle treatment to obtain a good powder statistics. Square and circle symbols represent compression and decompression respectively. The color lines are guides to eyes. Pressure induced phase transitions were observed at increasing pressures around 0.8 GPa, 5.2 GPa and 10 GPa. The absence of SHG signal indicates phase transformation into a centrosymmetric structure at 10 GPa.

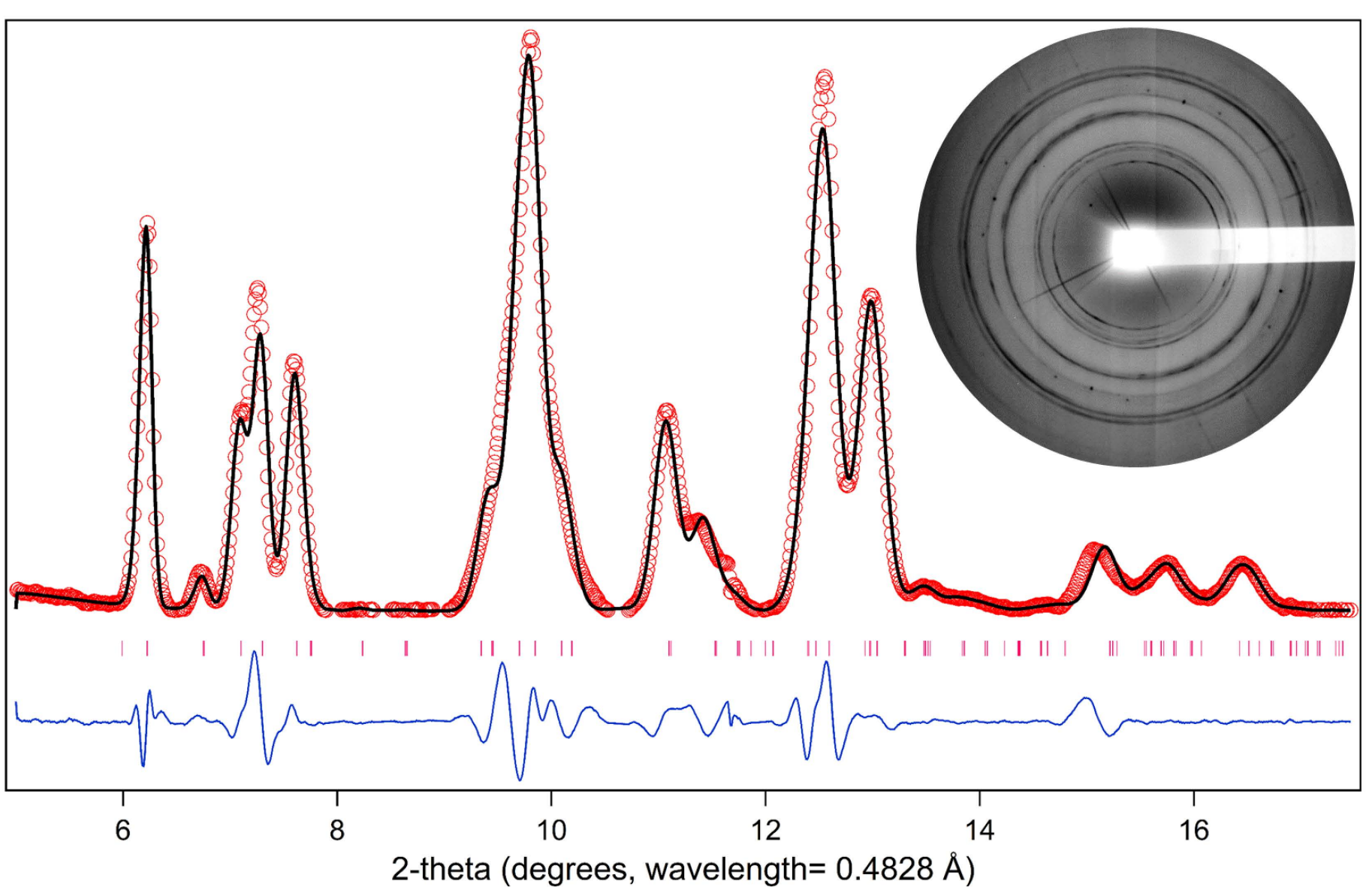

**Fig. 2** (a) Pawley analysis of the powder diffraction data of phase V ($P2_1/m$) urea at 11.9 GPa. The inset image plate shows the 2D diffraction pattern with good powder averaging statistics.

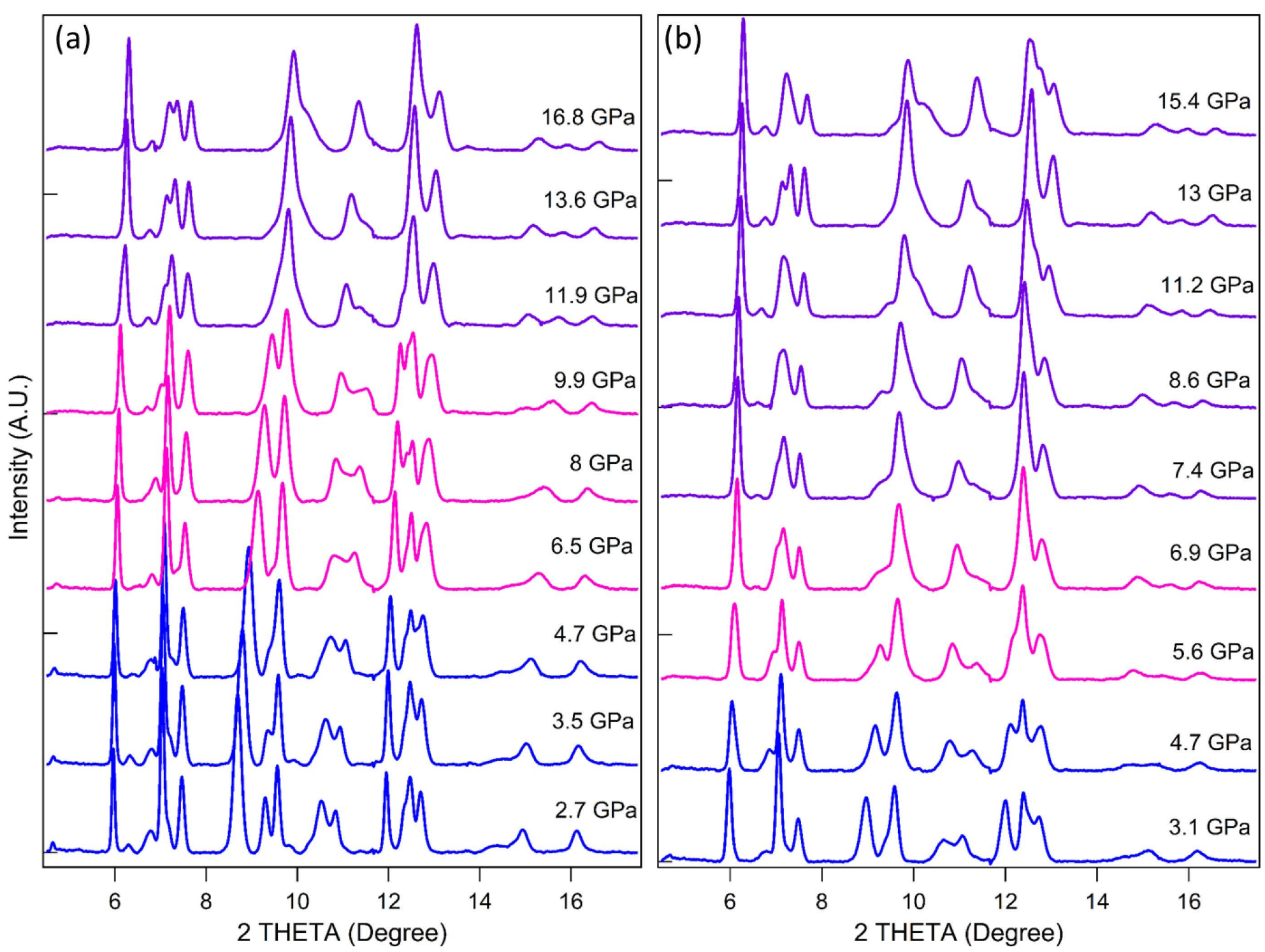

**Fig. 3** (a) XRD patterns of urea at selected pressures during compression and (b) decompression. The diffraction patterns are color coded. Blue curves represent Phase III ($P2_12_12_1$), pink curves represent phase X ($Pnnm$) , and purple curves represent phase V ($Pmna$)

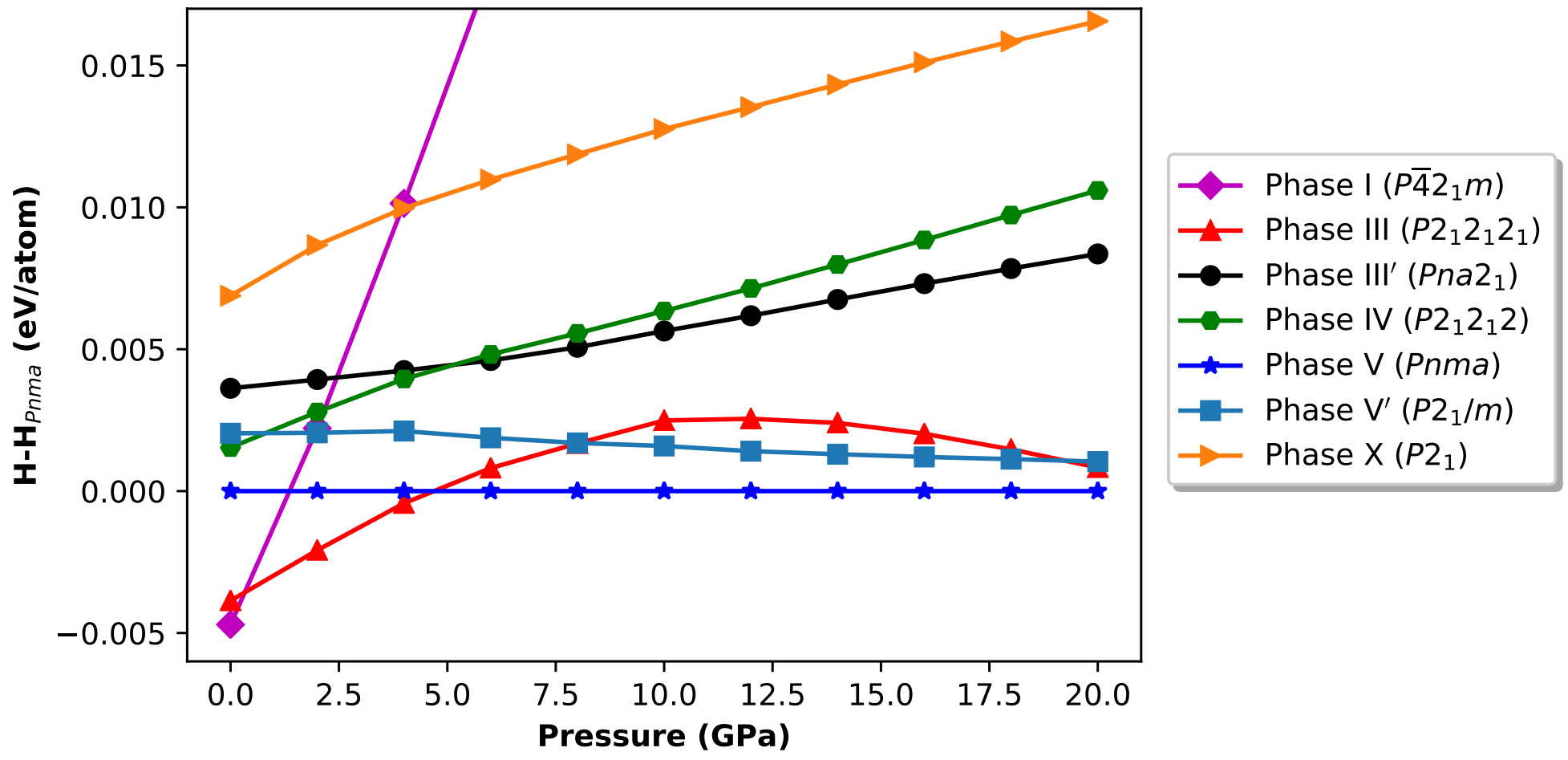

**Fig. 4** Calculated relative enthalpy difference of the predicted urea stable and meta stable polymorphs w.r.t phase V as a function of pressure.

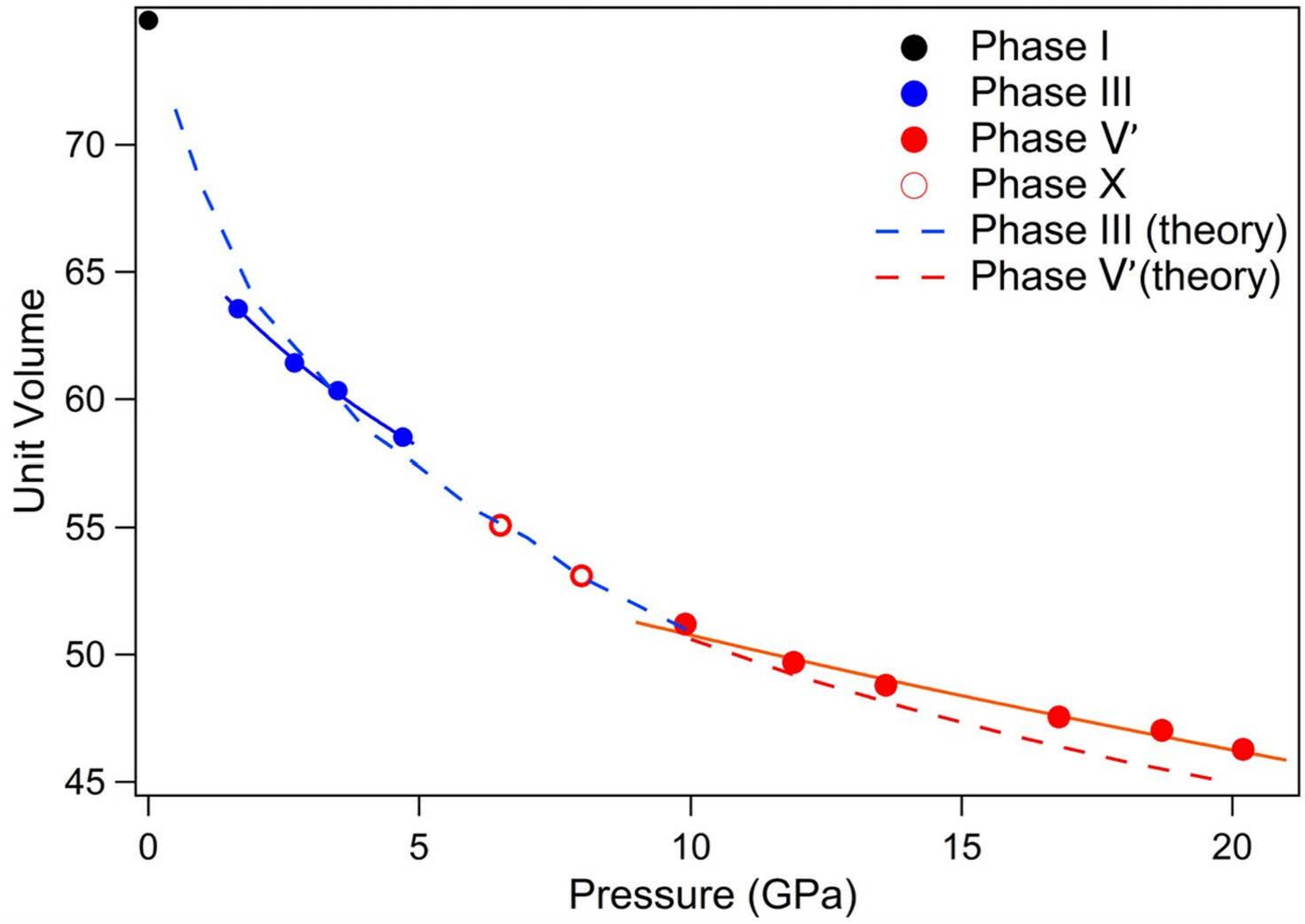

**Fig. 5** Unit volume per urea molecule as a function of pressure and the fitted equation of state, indicated by solid lines. The theoretical predicted equation of state is plotted in dashed lines and overlaid with the experimental results. The empty circles represent the unit volume of phase X, which exists between 6 to 10 GPa, and is likely a disorder phase result from the quantum melting of the hydrogen-bonding lattice that cannot be described by USPEX theory.

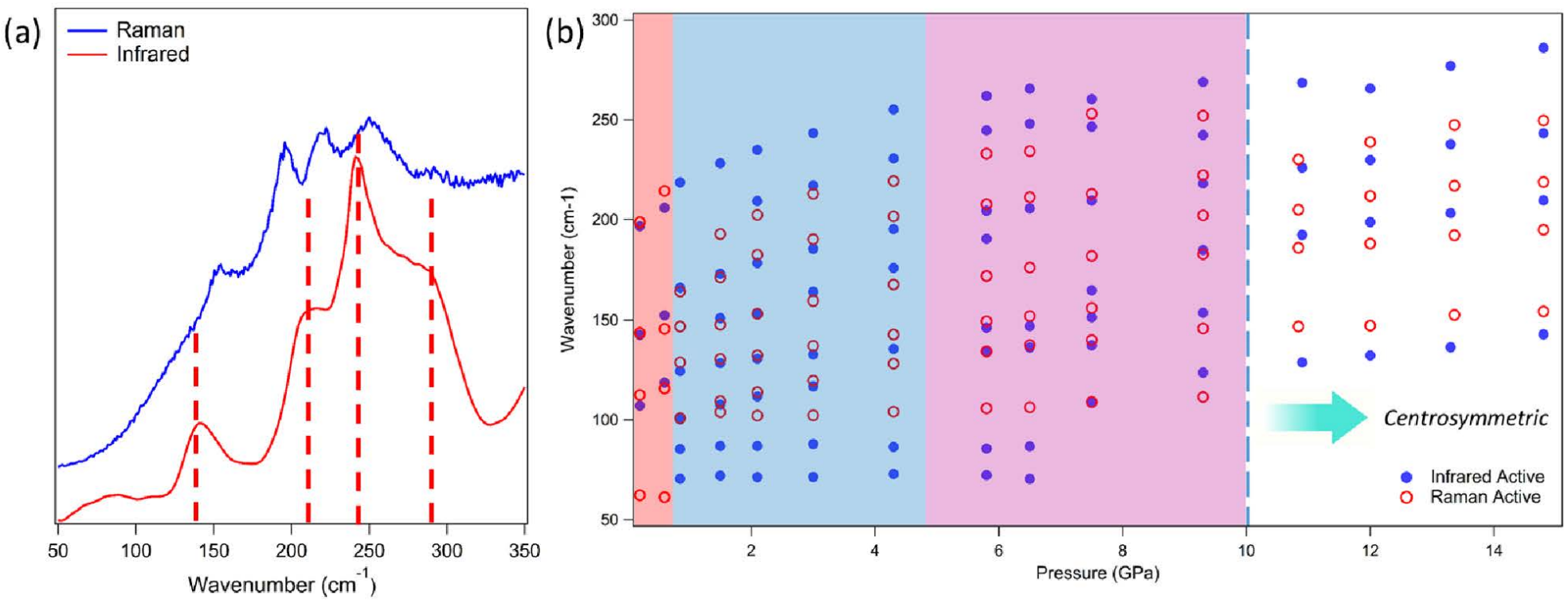

**Fig. 6** (a) The Raman and far infrared absorption spectra of urea phase $V'$ at 10 GPa. (b) Raman and infrared active lattice modes of high-pressure urea phases. The phase regions are color coded as phase I ($P42_11m$; red), phase III ($P2_12_12_1$; blue), phase X (Pnnm; purple), and phase $V'$ ($P2_1/m$; Z=6) beyond 10.0 GPa. The selection rules of the lattice phonons become mutually exclusive beyond 10 GPa which confirms the phase transformation from a non-centrosymmetric structure into a centrosymmetric structure.